\documentclass[]{aa}
\usepackage{graphicx}
\usepackage{natbib}
\usepackage{url}
\bibpunct{(}{)}{;}{a}{}{,}

\begin{document}

\title{Fast Spectral Fitting of Hard X-Ray Bremsstrahlung from Truncated Power-Law Electron Spectra}
\author{J.C. Brown\inst{1} \and J. Ka\v{s}parov\'{a}\inst{2} \and A.M. Massone\inst{3}
\and M.Piana\inst{4,3}}

\institute{Department of Physics and Astronomy, University of
Glasgow, Glasgow G12 8QQ, U.K.
\and Astronomick\'{y} \'{u}stav AV \v{C}R, v.v.i., Fri\v{c}ova 298, 25165 Ond\v{r}ejov, Czech Republic
\and CNR-INFM LAMIA, Via Dodecaneso 33, I-16146 Genova, Italy
\and Dipartimento di Informatica, Universit\`{a} di Verona, Ca' Vignal 2, Strada le Grazie 15, I-37134 Verona, Italy }

\offprints{J.C. Brown, \email{john@astro.gla.ac.uk}}

\authorrunning{Brown et. al.}
\titlerunning{Fast Spectral Fitting}
\date{Received/Accepted}

\abstract {Hard X-Ray bremsstrahlung continuum spectra, such as from
solar flares, are commonly described in terms of power-law fits,
either to the photon spectra themselves or to the electron spectra
responsible for them. In applications various approximate relations
between electron and photon spectral indices are often used for
energies both above and below electron low-energy cutoffs.} {We examine the form of the exact relationships in various
situations, and for various cross-sections, showing that empirical
relations sometimes used can be highly misleading and consider how
to improve fitting procedures.}
 {We obtain expressions for photon
spectra from single, double and truncated power-law electron spectra
for a variety of cross-sections and for the thin and thick target
models and simple analytic expressions for the Bethe-Heitler
cases.} {We show that above a low-energy cutoff the
Kramers and Bethe-Heitler results match reasonably well with results for exact
cross-sections up to energies around $100$ keV; that below the low-energy cutoff, Kramers and other constant spectral
index forms commonly used are very poor approximations to accurate results; but that our analytical forms are a
very good match.} {Analytical forms of the Bethe-Heitler photon spectra from general power-law electron spectra are
an excellent match to exact results for both thin and thick targets
and they enable much faster spectral fitting than evaluation of the full spectral integrations. }


\maketitle

\section{Introduction} \label{sect:intro}
Hard X-ray bremsstrahlung spectra are important diagnostics of flare
electron acceleration and propagation - e.g.,
\citet{1971SoPh...18..489B,1987ApJ...312..462L,1992SoPh..137..121J,1992A&A...265..278T,1994A&A...288..949P,2003ApJ...595L..97H}.
Extensive use of this diagnostic power has been enabled by the high
resolution spectra being observed by RHESSI
\citep{2002SoPh..210....3L}, which handles very large spectral and
dynamic ranges.  At low energies (a few keV for microevents and up
to 20 or so keV for large flares), the spectrum is usually
consistent with isothermal bresmsstrahlung - e.g.
\citep{2003ApJ...595L..97H} - while at higher energies it is usually
consistent with bremsstrahlung from (a sometimes broken) power-law
electron spectrum with a low-energy cutoff. One should bear in mind
that other, broadly similar, forms (e.g. shifted power-laws
$(E+E_*)^{-a}$) are also consistent with the data and that the data
$I(\epsilon)$ to be considered are, in general, those after
application of corrections for the albedo spectrum contribution
\citep{2006A&A...446.1157K} as well as instrumental effects.
However, here we focus on the properties of truncated power-law fits
(including broken power-laws) since these are so widely used.

We consider here mainly the properties of the higher energy
component and discuss the relationship between the 'local' spectral
indices $\delta(E)$ of the
source electrons and $\gamma(\epsilon)$ of the observed photons. In
data analysis it is quite common \citep[e.g.][]{2007arXiv0712.2544H} to assume
constant values of $\gamma(\epsilon)$ and of $\delta(E)$ over
specific finite energy ranges and definite linear relationships
between $\gamma$ and $\delta$ in these -  for example, in the energy
ranges below and above the electron cutoff. However, as we
show below, for general bremsstrahlung cross-sections, most such
relationships are at best approximate and the various ad hoc
relationships used can be quite misleading.  Our aim is to show the
exact form of these relationships for various cross-sections and
derive analytic expressions simple enough for easy use in fast
spectral analysis software.

For both the thin- and thick-target models this is accomplished by
first showing numerically that the Bethe-Heitler approximation for
the bremsstrahlung cross-section is reliable enough to reproduce
accurately the true photon spectrum corresponding to a truncated
power-law electron spectrum. Then for both models we obtain exact
analytical expressions for $I(\epsilon)$ and $\gamma(\epsilon)$
based on the Bethe-Heitler cross-section. Finally these expressions
are used to best fit simulated and measured photon spectra and
determine electron spectrum parameters. The effectiveness of this
method is assessed by comparisons with two different fitting
approaches currently employed: (1) fitting with a numerical
expression for $I(\epsilon)$ obtained by full numerical integration
\citep{2003ApJ...595L..97H} of truncated
 power-law electron spectra $\bar F(E)$ and ${\cal{F}}_0(E_0)$ (thin and thick target, respectively); 
(2) fitting data on $I(\epsilon)$ with a broken power-law
 characterized by two distinct constant photon spectral indices, one
 above and one below a 'knee' energy \citep[e.g.][]{2007arXiv0712.2544H}.
 With respect to the first approach, we find that our new
method yields equally good $\chi^2$ values but with substantially
higher computational speed (much higher in the case of thick
targets). With respect to the second approach, our method is no
faster but much more accurate and meaningful. In fact the
$I(\epsilon)$ used in approach (2) is unphysical, corresponding to
no real $\bar F(E)$ so it may give excessive $\chi^2$ values
due to large residuals near the knee.

Section 2 provides the general equations for bremsstrahlung spectra
for both collisionally thin and thick target sources. Section 3
establishes our notation for a power-law $\bar F(E)$ of constant
$\delta$ truncated below the low-energy cutoff $E_1$, and shows how arbitrarily truncated
and multiple (broken) power-laws can be expressed in terms of these.
Section 4 defines the Kramers and Bethe-Heitler approximate
bremsstrahlung cross-sections and contrasts the results they give
for $I(\epsilon)$ and $\gamma(\epsilon)$ for power-law electron
spectra compared with that for the exact cross-section. In Sections
5 and 6 we obtain analytic expressions, in both thin- and thick-target cases, for $I(\epsilon)$ and
$\gamma(\epsilon)$. In Section 7 we report numerical tests of the
speed and accuracy of using these to fit real and simulated data, as
compared with other approximate methods and with full integration.
Section 8 summarises our conclusions.

\section{Thin and thick target bremsstrahlung and energy losses}
\hspace{6mm}
For a general inhomogeneous optically thin source of plasma density $n({\bf r})$ and electron flux energy spectrum $F(E, {\bf r})$ in volume $V$, the bremsstrahlung photon flux energy spectrum $I(\epsilon)$ (cm$^{-2}$s$^{-1}$ per unit $\epsilon$ at Earth distance R) can be written \citep{1971SoPh...18..489B}:
\begin{equation}\label{ThinI}
I(\epsilon)=\frac{\bar n V}{4\pi R^2}  \int_\epsilon^{\infty} \bar
F(E)Q(\epsilon,E) dE~~~,
\end{equation}
with
\begin{equation}\label{density}
\bar n= \int_V n dV/V
\end{equation}
and
\begin{equation}\label{fbar}
\bar F(E) =\int_V n({\bf
r})F(E,{\bf r})dV/(\bar nV)
\end{equation}
where $Q(\epsilon,E)$ is the bremsstrahlung cross-section differential in photon energy $\epsilon$. In general, $\bar F(E)$ and
$Q(\epsilon,E)$ have to be treated as anisotropic and Equation (\ref{ThinI}) involves an integral over solid angle
\citep{1972SoPh...26..441B,2004ApJ...613.1233M} though most data treatments assume source isotropy.

In a purely collisional thick target, ${\bar{F}}(E)$ is related to the thick-target injection rate
spectrum ${\cal{F}}_0(E_0)$
(electrons per second per unit injection energy $E_0$) through equation
\begin{equation}\label{FbarforFo}
\bar F(E)=\frac{1}{K\bar nV}E\int_E^\infty {\cal F}_o(E_o)dE_o
\end{equation}
regardless of $Q(\epsilon,E)$, with $K=2\pi e^4 \Lambda$ and
$\Lambda$ the Coulomb logarithm - \citet{1988ApJ...331..554B}.
Though they consider only the case $K=2\pi e^4 \Lambda$ for
collisional losses only in a uniformly ionized target, Equation
(\ref{FbarforFo}) applies to any energy loss rate coefficient $K(E)$
if the loss rate can be written in the form  $dE/dN = - K(E)/E$
 where $N$ is the column density along the electron path such as for collisional
 losses at high $E$ with relativistic correction. Synchrotron
 losses cannot be written in this way unless the magnetic field and electron pitch angle distribution do not
 vary along the path, but these also only matter at high energies. The most serious approximation involved in using constant $K$ for lower energies is 
in (common) neglect of the fact that $K$ varies
  with target hydrogen ionization $x$ \citep{1973SoPh...29..421B,1978ApJ...224..241E}
   being $K=2\pi e^4 \Lambda(1+ax)/(1+a)$ with $a\approx 1.6$. This is important around
   the energies of electrons $E_*\approx(2KN_*)^{1/2}$ stopping around the flare transition
   zone depth $N_*$. Apart from early stages of flares, prior to much evaporation, $E_*$ is
    well above typical values of $E_1$ considered here. Since we are concerned mainly with
  small $E$ around low-energy cutoff values, henceforth we address only the case of constant $K$
and take $\Lambda=25$.

\section{Single and broken power-laws}
\hspace{6mm} We consider first the widely used single power-law form
with low-energy cutoff for the thin target $\bar F(E)$
\begin{equation}\label{PowerlawFbar}
\bar F(E)=(\delta-1) \frac{F_1}{E_1}\left\{ \begin{array}{ll}
\left[\frac{E}{E_1}\right]^{-\delta} & E\ge E_1 \\
\hspace{0.3cm}0 & E <E_1
\end{array}\right.
\end{equation}
where $F_1=\int_{E_1}^\infty \bar F(E)dE$ is the total mean electron flux at $E\ge E_1$, a low-energy cutoff\footnote{If one wishes to use an electron flux reference energy $E_*$ distinct from the cutoff energy $E_1$ one must replace $F_1$ in Eq. (\ref{PowerlawFbar}) by $F_*(E_*/E_1)^{-\delta+1}$.}, and $\delta$ is the (thin target; \citet{1971SoPh...18..489B}) constant electron spectral index.

Before considering the photon spectral properties of such single
power-law electron spectra we note that our results for these can
easily be generalised to fitting of double (broken) power-laws in
${\bar{F}}(E)$ with lower and upper cutoff energies. The following
decomposition expressions apply equally well to any broken and
truncated power-laws such as in photon space $I(\epsilon)$. The
general case is
\begin{equation}\label{a}
{\bar F}(E)= \left\{ \begin{array}{lr} 0 & E < E_c \nonumber \\
AE^{-\delta_1} & E_c \leq E < E_b \nonumber \\
AE_b^{-\delta_1+\delta_2}E^{-\delta_2} & E_b \le E < E_a \nonumber \\
0 & E \geq E_a
\end{array}
\right.
\end{equation}
where $A$, $E_c$, $E_b$, $E_a$, $\delta_1$, $\delta_2$ are constants. Writing the parametrized single power-law $F_{pl}(c,d,E^*)$ as
\begin{equation}\label{fpl}
F_{pl}(c,d,E^*) = \left\{ \begin{array}{lr} cE^{-d} & E \geq E^* \nonumber \\
0 & {\mbox{otherwise}}~~,
\end{array}
\right.
\end{equation}
${\bar{F}}(E)$ in (\ref{a}) can be always written as
\begin{eqnarray}\label{fbar-truncated}
\bar F(E)& = & F_{pl}(A,\delta_1,E_c)-F_{pl}(A,\delta_1,E_b)+  \\
& & F_{pl}(AE_{b}^{-\delta_1+\delta_2},\delta_2,E_b)- F_{pl}(AE_{b}^{-\delta_1+\delta_2},\delta_2,E_a) \nonumber.
\end{eqnarray}
Hence the corresponding $I(\epsilon)$ can be found simply as the sum and difference of the relevant $I(\epsilon)$ expressions for single power-laws.

For the thick target model, we have to revisit the problem and evaluate the form of $\bar F(E)$ and hence of
$I(\epsilon)$ for a truncated power law in ${\cal F}_0(E_0)$, not in $\bar F(E)$. Note also that we have to distinguish between the spectral index $\delta_o$ for a pure power-law injection spectrum ${\cal F}_0(E_0)$
from the index $\delta$ for $\bar F(E)$. For an injection spectrum truncated at $E_0\le E_{01}$
\begin{equation}
\label{PLawFo}
{\cal F}_0(E_0)  = \left\{ \begin{array}{ll}
(\delta_0-1)\frac{{\cal F}_{01}}{E_{01}}\left[\frac{E_0}{E_{01}}\right]^{-\delta_0} & E_0 \ge E_{01} \\
0 & E_0 < E_{01}~,
\end{array}\right.
\end{equation}
where ${\cal{F}}_{01}=\int {\cal{F}}(E_0) dE_0$ and equation (\ref{FbarforFo}) gives
\begin{equation}
\label{FbarThick}
\bar F(E)= \frac{{\cal F}_{01}E_{01}}{K\bar n V} \left\{ \begin{array}{ll}
\left[\frac{E}{E_{01}}\right]^{-\delta_0+2} & E\ge E_{01} \\
\hspace{0.2cm}\frac{E}{E_{01}} & E<E_{01}~,
\end{array}\right.
\end{equation}
i.e., the relation between $\delta_0$ and $\delta$ at $E\ge E_{01}$
being $\delta = \delta_0 -2$. To find $\bar F(E)$ from a broken and truncated power-law form of
the injected thick target ${\cal F}_0(E_0)$, one would first use the analogy of 
expression (\ref{fbar-truncated}) to get the total ${\cal F}_0(E_0)$ as the sum
of a set of single power-law ${\cal F}_0(E_0)$ forms and use expression (\ref{FbarThick}) for each term in the sum to get the corresponding $\bar F(E)$.

\section{Cross-sections} \hspace{6mm}
The bremsstrahlung
cross-section $Q(\epsilon,E)$ can be written as
\begin{equation}\label{cross}
Q(\epsilon,E) = \frac{Q_o m c^2}{\epsilon E} q(\epsilon,E)~,
\end{equation}
incorporating a high $Z$ element correction factor $\sum_{Z} A_Z Z^2$ in $Q_o$, the Gaunt factor $q(\epsilon,E)$ depending on the
actual cross-section used. For the Kramers approximation
\begin{equation}\label{kramers}
q(\epsilon,E) = q_K(\epsilon,E) =1~;
\end{equation}
while for the non-relativistic Bethe-Heitler approximation
\begin{equation}\label{bethe-heitler}
q(\epsilon,E) = q_{BH}(\epsilon,E) =\log\frac{1+\sqrt{1-\epsilon/E}}{1-\sqrt{1-\epsilon/E}}~.
\end{equation}
The most widely used isotropic formula (neglecting electron-electron
bremsstrahlung which is important at high energies
\citep{2007ApJ...670..857K}) is $q(\epsilon,E) =
q_{3BN}(\epsilon,E)$ corresponding to the 3BN formula from
\citet{1959KochMotz} used in inversion and forward fits in
\citet{2003ApJ...595L.127P,2005SoPh..226..317K,2006ApJ...643..523B,1992SoPh..137..121J}.
In the case of a truncated power-law, the sensitivity of the
predicted thin- and thick-target photon spectra to the form of the
cross-section used is illustrated in Figure 1 by numerically
computing $\gamma(\epsilon)$ for $\delta=3$, $\delta=5$;
$\delta_0=5$, $\delta_0=7$ and various $q$. It is clear firstly
that, particularly below the cutoff, Bethe-Heitler results are a
rather good approximation for computing local photon spectral
indices $\gamma(\epsilon)$. Second, above and especially below the
cutoff $\gamma(\epsilon)$ is not in any way constant, as sometimes
assumed \citep{2007arXiv0712.2544H}. Figure 1 thus clearly suggests
that the analytical expressions of $I(\epsilon)$ and
$\gamma(\epsilon)$ based on  the Bethe-Heitler formula could be
useful for spectral computation in forward spectral fitting.

Note that Figure 1 shows results for only one value of $E_1$ ($E_{01}$) and
presents them only as functions of $\epsilon/E_1$ ($\epsilon/E_{01}$) whereas one might
in general expect results to depend on $\epsilon$ and $E_1$ ($E_{01}$)
separately. However, we carried out a range of test calculations for
several different $E_1$ ($E_{01}$) in the few deka-keV range and found that
results were, to a very good approximation, functions only of
$\epsilon/E_1$ ($\epsilon/E_{01}$). Secondly, the most general $Q$ is actually
anisotropic and the above expressions have to be generalised to
integrate over electron angle as well as energy \citep{2004ApJ...613.1233M}.
However, the effect of this on $I(\epsilon)$ is small at low
energies and in any case is mainly an $\epsilon$-independent scaling
rather than a spectral effect. In fact some numerical experiments
showed that there is little effect on our conclusions of using
anisotropic ${\bar F}, Q$.

%\begin{figure}
%\begin{tabular}{c}
%  \resizebox{\hsize}{!}{\includegraphics{fig1a}}
%   \\
%(a) \\
%  \resizebox{\hsize}{!}{\includegraphics{fig1b}} \\
%(b)
%\end{tabular}
%  \caption{TO BE DONE}
%  \label{<Your label>}
%\end{figure}

\begin{figure*}
\centering
\includegraphics[width=8cm]{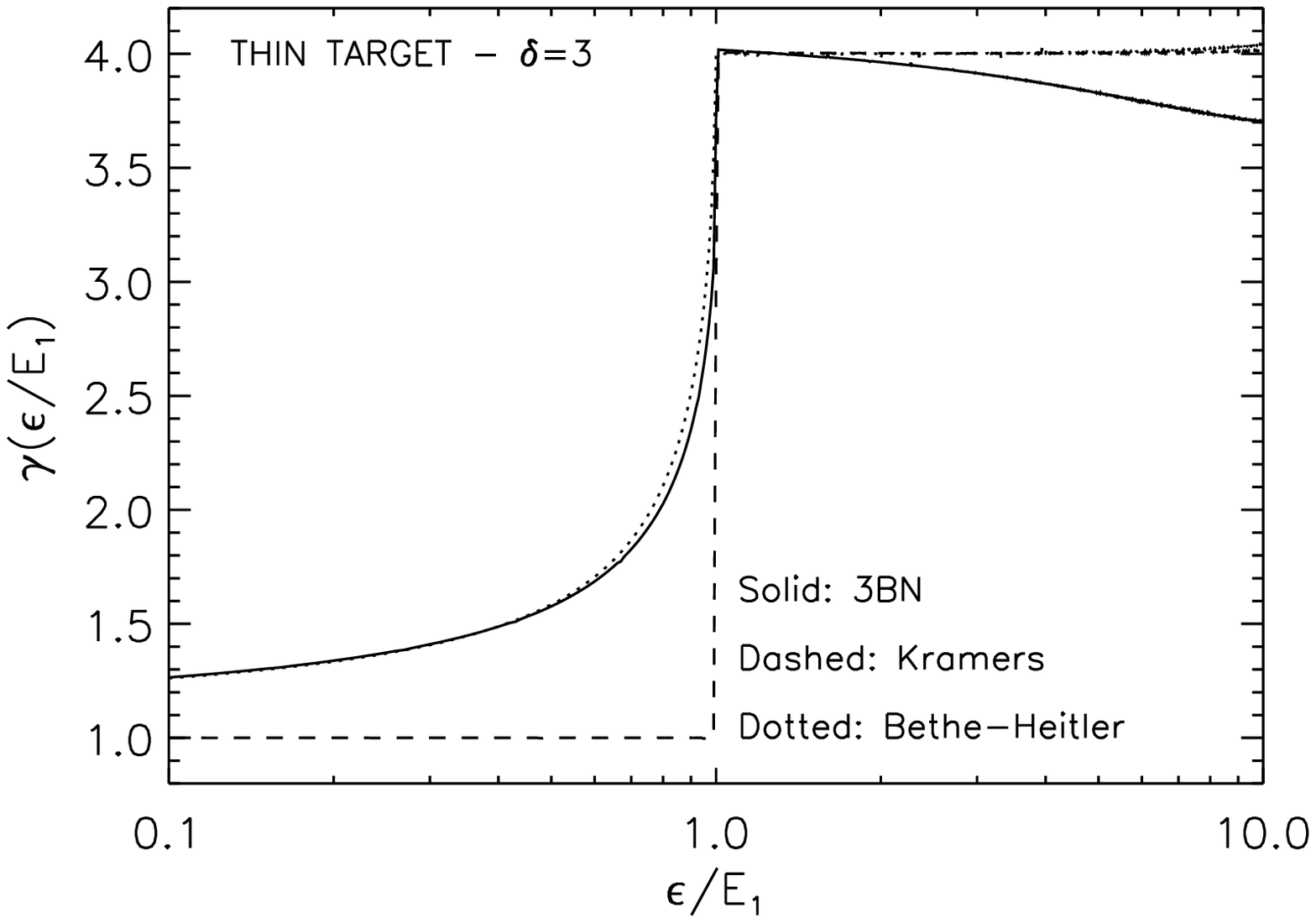}
\includegraphics[width=8cm]{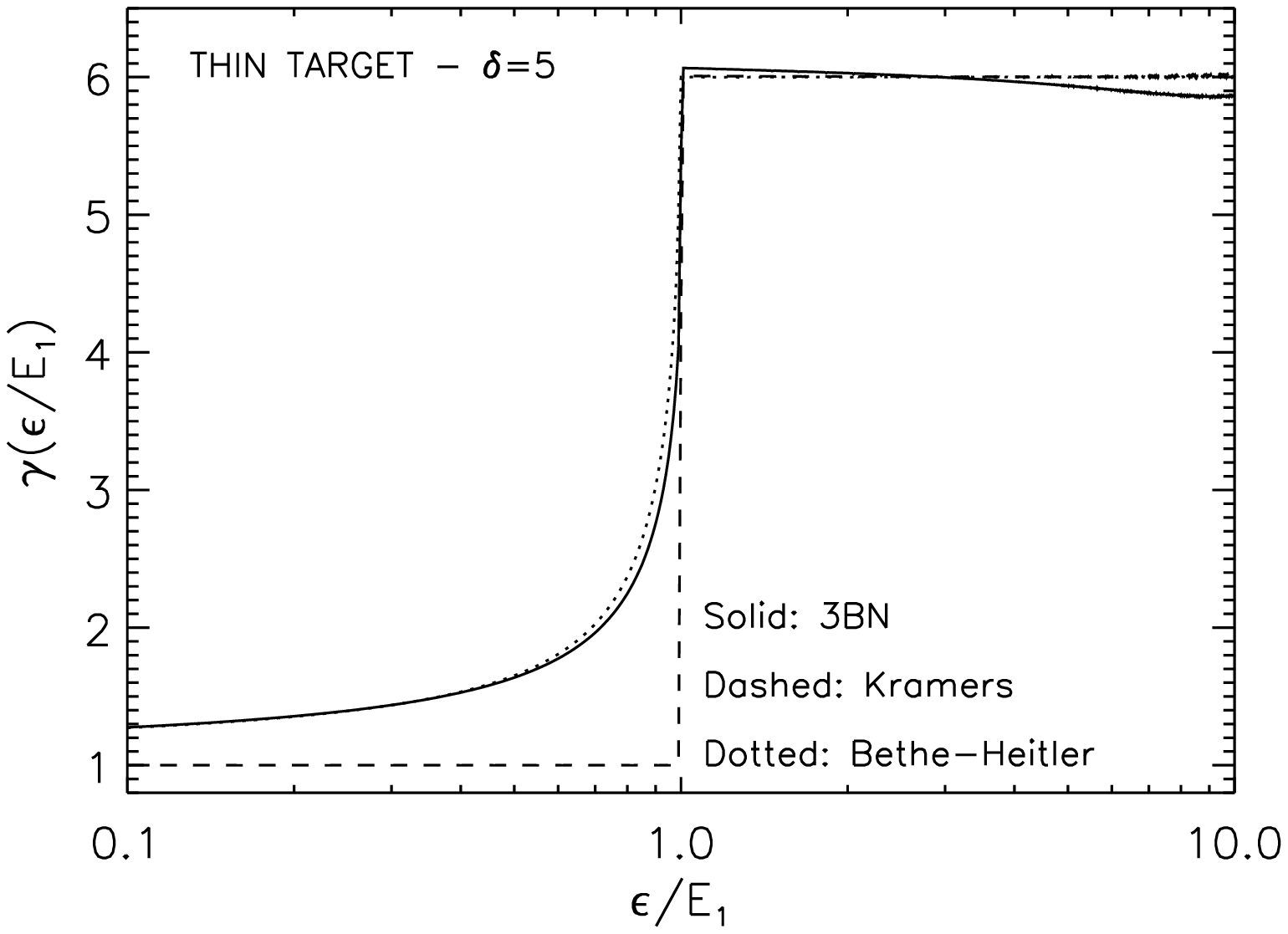}

\hspace{1cm}(a)\hspace{8cm}(b)

\includegraphics[width=8cm]{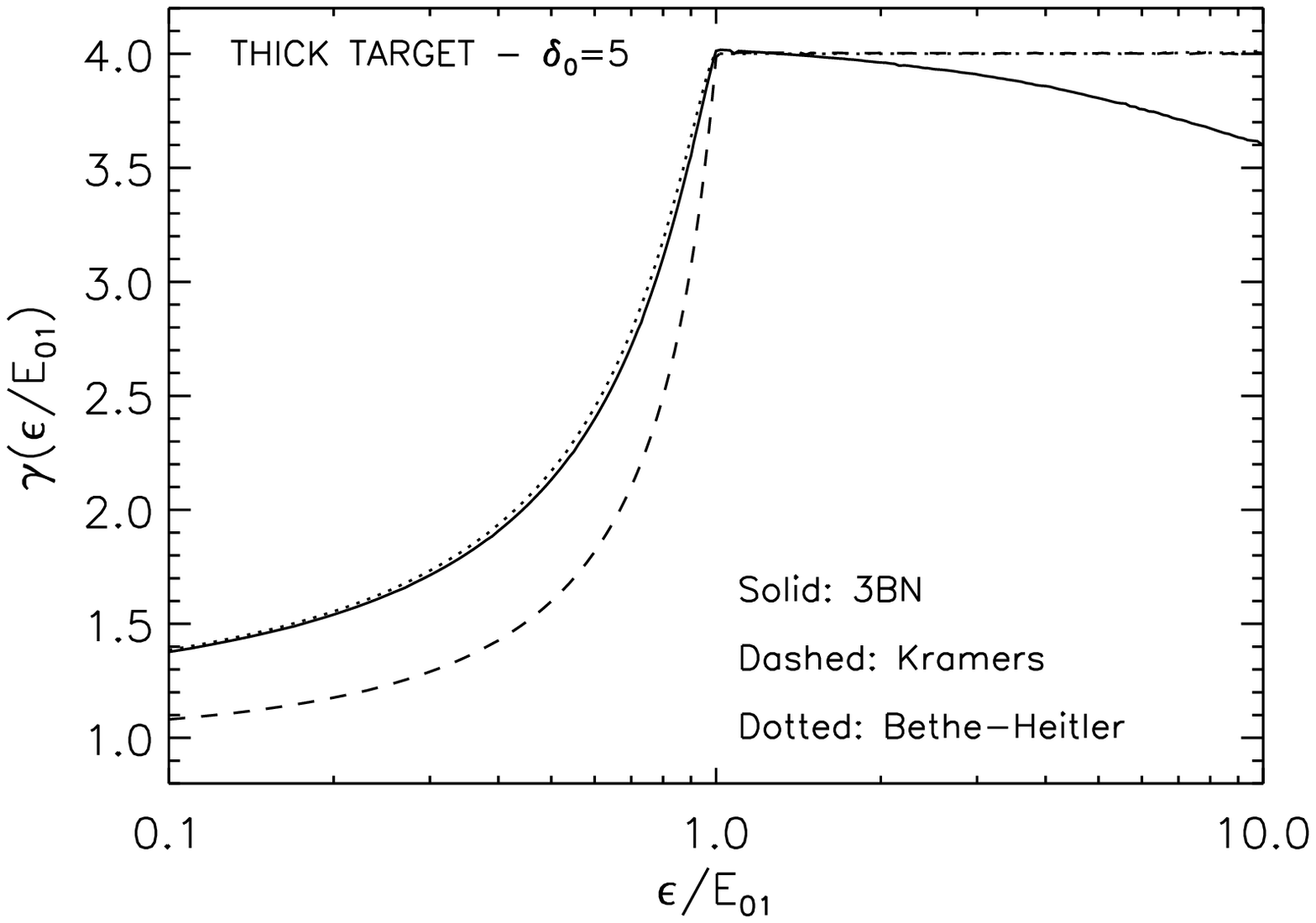}
\includegraphics[width=8cm]{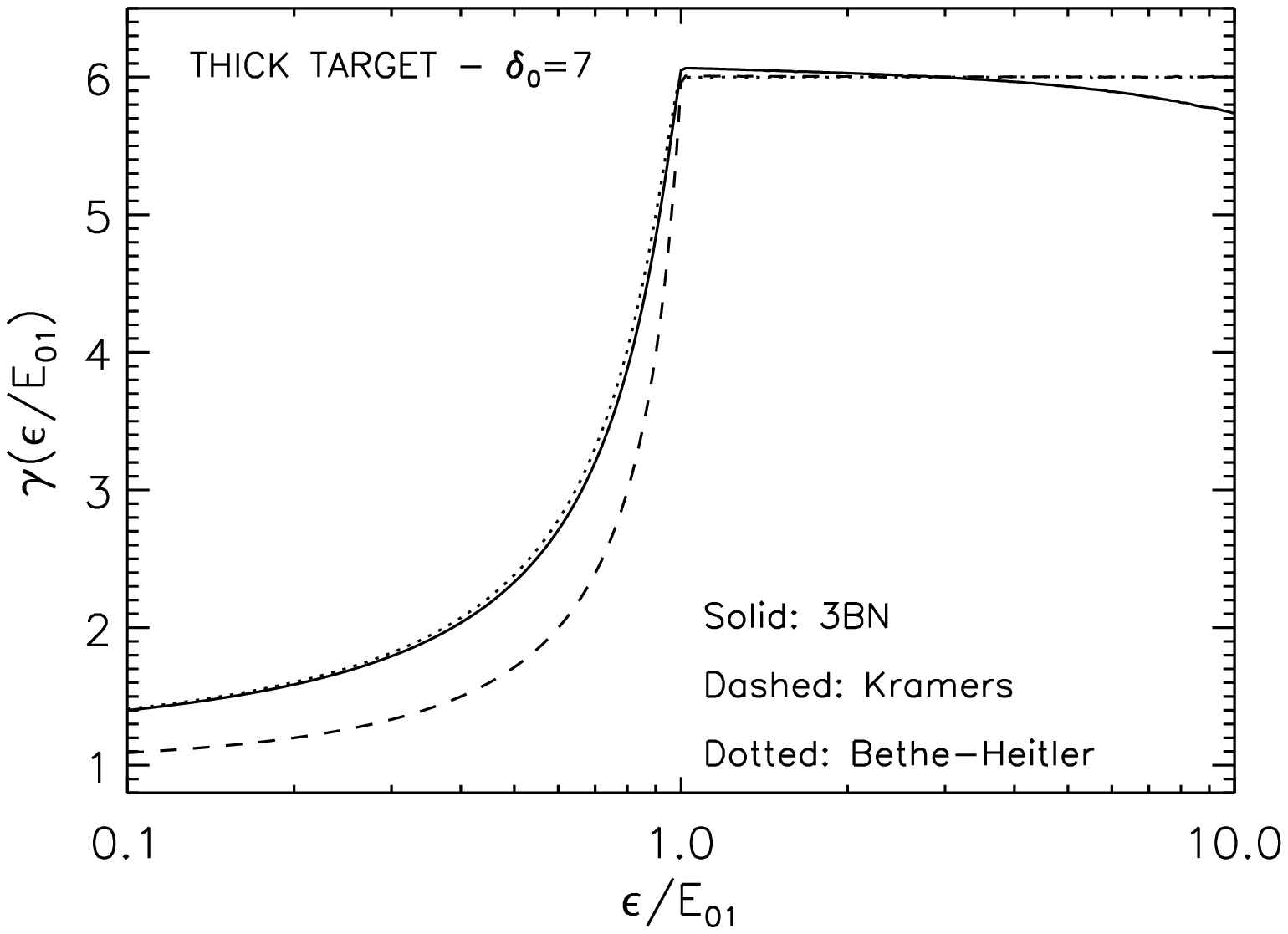}

\hspace{1cm}(c)\hspace{8cm}(d) \caption{Local bremsstrahlung photon
spectral index $\gamma(\epsilon)$ behavior computed numerically for
three different $q(\epsilon,E)$ - Kramers, Bethe-Heitler and 3BN.
(a) Thin target with $\delta=3$. (b) Thin target with $\delta=5$.
(c) Thick target with $\delta_0=5$. (d) Thick target with
$\delta_0=7$.} \label{fig1}
\end{figure*}

\section{Thin target spectra and spectral indices}
\hspace{6mm} Using Equation (\ref{cross}), Equation (\ref{ThinI}) becomes
\begin{equation}\label{ThinI-new}
I(\epsilon) = \frac{Q_0 mc^2}{4\pi R^2} \frac{{\overline{n}}V}{\epsilon} \int_{\epsilon}^{\infty}
{\bar{F}}(E) q(\epsilon,E) \frac{dE}{E}~.
\end{equation}
We are interested here in results for photon spectra $I(\epsilon)$
when ${\bar{F}}(E)$ is a single power-law with constant $\delta$ and
a low-energy cutoff (Equation (\ref{PowerlawFbar})), for various
forms of $q(\epsilon, E)$ considered, and also on the form of local
photon spectral index defined by (Brown and Emslie 1988)
\begin{equation}
\label{Defgamma(eps)} \gamma(\epsilon)=
-\frac{\epsilon}{I}\frac{dI}{d\epsilon}=-\frac{d\log I}{d\log \epsilon}~.
\end{equation}
for comparison with the use \citep[e.g.][]{2007arXiv0712.2544H} of
constant $\gamma$ approximations. For any $q(\epsilon,E)$ equation
(\ref{ThinI-new}) can be written
\begin{equation}\label{PowerlawIGen}
I(\epsilon) =
\frac{\delta-1}{\delta}C\frac{E_1}{\epsilon}\int_0^{min[1,(E_1/\epsilon)^\delta]}q(\epsilon,E_1/x^{1/\delta})dx
\end{equation}
where $x=(E_1/E)^\delta$, and
\begin{equation}
C=\frac{Q_0 mc^2}{4\pi R^2} \frac{{\overline{n}}VF_1}{E_1^2}
\end{equation}

Note that in special cases where $q(\epsilon,E) = q(\epsilon/E)$
only, $\gamma(\epsilon)$ takes the form
\begin{equation}\label{gammaSEP}
\gamma(\epsilon)= \left\{ \begin{array}{ll}
\delta+1 & \epsilon \ge E_1 \\
1-\frac{d}{d\log \epsilon}\log\left[\int_0^1 q(x^{1/\delta}\epsilon/E_1)\right] dx & \epsilon < E_1~.
\end{array}
\right.
\end{equation}

\subsection{Kramers cross-section}\label{sec:KramersThin}
For Kramers $q(\epsilon,E)=1$ and we have immediately
\begin{equation}
\label{IKramersIeps}
I(\epsilon) = \frac{\delta-1}{\delta} C \frac{E_1}{\epsilon}\left\{ \begin{array}{ll}
\left[\frac{E_1}{\epsilon}\right]^{\delta} & \epsilon \ge E_1 \\
\hspace{0.3cm}1 & \epsilon < E_1
\end{array}\right.
\end{equation}
and
\begin{equation}
\label{gammaKramers(eps)} \gamma(\epsilon)=\left\{ \begin{array}{ll}
\delta+1 & \epsilon\ge E_1 \nonumber\\
\hspace{0.3cm}1 & \epsilon < E_1~.
\end{array}\right.
\end{equation}

\subsection{Bethe-Heitler cross-section}
\hspace{6mm} The spectrum and the spectral index can be written
solely in terms of the electron spectral index $\delta$ and the
dimensionless parameter $a=\epsilon/E_1$. Integration by parts leads
to:

\begin{eqnarray}
\label{bhI}
I(\delta,a)&=&\frac{\delta-1}{\delta} C
\left\{ \begin{array}{lr}
\frac{1}{a^{\delta+1}} B(\delta,\frac{1}{2})   &    a \ge 1   \\
& \\
\frac{1}{a}\left[\log\frac{1+\sqrt{1-a}}{1-\sqrt{1-a}} + \frac{1}{a^\delta}~B_a(\delta,\frac{1}{2})\right] &  a<1
\end{array}\right.
\end{eqnarray}
with $B(\alpha,\beta) = \int_0^1 x^{\alpha-1} (1-x)^{\beta-1} dx$ and
$B_x(\alpha,\beta) = \int_0^x t^{\alpha-1} (1-t)^{\beta-1} dt$ the incomplete beta function.

For the spectral index:

\begin{equation}\label{bhgamma}
\gamma(\delta,a)=\left\{ \begin{array}{lr} 1+\delta & a \geq 1 \\
1+ \frac{\frac{\textstyle \delta}{\textstyle a^\delta}~\textstyle B_a(\delta,\frac{1}{2})}{\textstyle \log\left[\frac{\textstyle 1+\sqrt{1-a}}{\textstyle 1-\sqrt{1-a}}\right]+\frac{\textstyle 1}{\textstyle a^\delta}~B_a(\delta,\frac{1}{2})} & a < 1
\end{array}
\right.
\end{equation}
We note that the previous analytical formulas for $I(\delta,a)$ and
$\gamma(\delta,a)$ are written in terms of beta functions and
incomplete beta functions. The computation of these functions is
included in standard library routines for data visualization and
analysis, making the implementation of the exact formulas
(\ref{bhI}) and (\ref{bhgamma}) easy and fast.

\section{Thick target spectra and spectral indices}
\hspace{6mm}
Inserting (\ref{FbarThick}) into (\ref{ThinI-new}) leads to
\begin{equation}\label{Ithick-general}
I(\epsilon) = \frac{D}{\delta_0-2}\frac{E_{01}}{\epsilon}\times \left\{
\begin{array}{ll}
\int_{0}^{(\frac{E_{01}}{\epsilon})^{\delta_0-2}} q(\epsilon,E_{01}/y^{\frac{1}{\delta_0-2}})dy & \epsilon \geq E_{01}\\
&\\
\int_{0}^{1}q(\epsilon,E_{01}/y^{\frac{1}{\delta_0+2}})dy + \\ (\delta_0-2) \int_{\epsilon/E_{01}}^{1}q(\epsilon,\xi E_{01})d\xi &
\epsilon < E_{01} ~,
\end{array} \right.
\end{equation}
where $D=\frac{Q_0 m c^2{\cal F}_{01}}{4 \pi R^2 K}$, the integrals
on the right hand side of (\ref{Ithick-general}) being expressed
below in terms of the dimensionless parameter $a=\epsilon/E_{01}$, the
constant $D$ and the spectral index $\delta_0$ for the truncated
electron power-law ${\cal{F}}_0(E_0)$.

\subsection{Kramers cross-section}
\hspace{6mm} Integration by parts with Kramers unity Gaunt factor
gives
\begin{equation}
\label{ThickIKramers}
I(\delta_0,a)=  \frac{D}{\delta_0-2} \left\{ \begin{array}{ll}
a^{-\delta_0+1}  & a \ge 1 \\
\frac{1+(\delta_0 -2)(1-a)}{a} & a < 1
\end{array}\right.
\end{equation}
and
\begin{equation}
\label{ThickgammaKramers}
\gamma(\delta_0,a)=   \left\{ \begin{array}{ll}
\delta_0-1  & a \ge 1  \\
\frac{\delta_0-1}{1+(\delta_0-2)(1-a)}& a < 1~.
\end{array}\right.
\end{equation}

\subsection{Bethe-Heitler cross-section}
\hspace{6mm}
Integration by parts with the Bethe-Heitler Gaunt factor $q_{BH}$ leads to

\begin{eqnarray}
\label{ThickIBH}
I(\delta_0,a)&=&
\frac{D}{\delta_0-2}\frac{1}{a} \left\{ \begin{array}{ll}
B(\delta_0-2,\frac{1}{2})\left[\frac{1}{a}\right]^{\delta_0-2}   &   \hspace{-0.55cm} a \ge 1   \\
&\\
\left[\delta_0-1-\frac{\delta_0-2}{2}a \right]\log\left[\frac{1+\sqrt{1-a}}{1-\sqrt{1-a}}\right]+ & \\
\hspace{0.1cm} - (\delta_0-2)\sqrt{1-a} + & \\
\hspace{0.1cm} + \frac{1}{a^{\delta_0-2}}B_a(\delta_0-2,\frac{1}{2})&  \hspace{-0.55cm} a < 1
\end{array}\right.
\end{eqnarray}
and
\begin{equation}\label{ThickgammaBH}
\gamma(\delta_0,a)= \left\{ \begin{array}{ll}
\delta_0 - 1 & a \geq 1 \\
\frac{A_1 + A_2 B_a(\delta_0-2,\frac{1}{2})}{A_3 - \sqrt{1-a} + A_4 B_a(\delta_0-2,\frac{1}{2})} & a<1,
\end{array}
\right.
\end{equation}
where
\begin{equation}\label{A}
A_1 = A_1(\delta_0,a)=\frac{\delta_0-1}{\delta_0-2}\log\frac{1+\sqrt{1-a}}{1-\sqrt{1-a}}~~,
\end{equation}
\begin{equation}\label{B}
A_2 = A_2(\delta_0,a)=\frac{\delta_0-1}{\delta_0-2} \frac{1}{a^{\delta_0-2}}~~,
\end{equation}
\begin{equation}\label{C}
A_3 = A_3(\delta_0,a)=\left(\frac{\delta_0-1}{\delta_0-2}-\frac{a}{2} \right)\log\frac{1+\sqrt{1-a}}{1-\sqrt{1-a}}
\end{equation}
and
\begin{equation}\label{D}
A_4 = A_4(\delta_0,a)=\frac{1}{\delta_0-2}\frac{1}{a^{\delta_0-2}}~~.
\end{equation}

Thus, as in the thin target case, integral expressions for
$I(\delta_0,a)$ and $\gamma(\delta_0,a)$ can be written analytically in
terms of beta functions and incomplete beta functions.

\section{Tests against data and comparison with other fitting methods}

To show the usefulness of our formulation in terms of accuracy and
speed, we have tried it out on simulated and real data. Our specific
goals here for simulated noisy data are to determine :

\begin{enumerate}
\item How good and how fast is use of our functional approximations to $I(\epsilon)$ in fitting
data to estimate [$\delta,{\overline{n}}V F_1, E_1$] or
[$\delta_0,{\cal{F}}_{01}, E_{01}$] for a [thin] or [thick] target single
truncated power-law [$\bar{F}(E)$], or [${\cal F}_0(E_0)$], in
comparison with other fitting routines. To answer this we use
simulated data $I(\epsilon)$ generated with NASA SolarSoft (SSW)
routines using the full bremsstrahlung cross-section and numerical
integration over $E$ or $E_0$, respectively
\citep{2003ApJ...595L..97H}.
\item How well do the speed and accuracy compare with other approaches? In particular with :
(a) \citet{2003ApJ...595L..97H} who, in carrying out the best fit
parameter searches, perform full integrations with the full cross-section in each iterative step; and (b) an approach which, 
instead of fitting the $I(\epsilon)$ predicted for an electron power law with cutoff, fit a parametric piecewise
power-law photon $I(\epsilon)$ with distinct constant photon
spectral indices $\gamma_1,\gamma_2$ at $\epsilon\le,\ge
\epsilon_b$, with $\gamma_1$ either a free fit parameter or
prescribed, e.g. as in \citet{2007arXiv0712.2544H} where $\gamma_1 = 1.5$ (to be compared with $\gamma(\epsilon)$ in Figure 1).
Loosely speaking, for example in the case of thin targets, the value of the
photon break energy $\epsilon_b$ is meant to reflect an electron
low-energy cutoff energy $E_1$, though in reality such a
photon spectrum does not correspond to any real (non-negative)
electron spectrum except for Kramers cross-section in which case
$\gamma_1 = 1$, $\gamma_2=\delta+1$.
\item How each of the above findings changes when we add a
reasonable isothermal contribution to $I(\epsilon)$ for
thermal parameters $EM$ and $T$.
\end{enumerate}

We have carried out these simulated data comparisons for the
following parameter sets: $\delta,\delta_0 = 3,5$, $E_1,E_{01}=10, 20, 30$~keV, $\bar n V F_1 =
5,20\times10^{55}$~electrons~cm${}^{-2}$~s${}^{-1}$; ${\cal F}_{01}=5,20\times10^{35}$~electrons~s${}^{-1}$,
$EM=0.5,1\times10^{49}$~cm${}^{-3}$, and $kT=1,1.5$~keV.

$I(\epsilon)$ were generated using SSW routines for thin and thick
target (\verb(f_thin.pro( and  \verb(f_thick.pro(, respectively) and optionally with isothermal component
(\verb(f_vth.pro() in the 3~--~100~keV energy range with 1~keV
energy binning. Then, these $I(\epsilon)$ were converted to counts
using the RHESSI detector response matrix for attenuator state 0.
Finally, we added Poisson noise. Such simulated count spectra were
then fitted within the
OSPEX environment\footnote{\url{http://hesperia.gsfc.nasa.gov/ssw/packages/spex/doc/ospex_explanation.htm}}
in the 3~--~100~keV range.

The results of comparisons for the simulated spectra {\it without} a
thermal component are as follows:

\begin{enumerate}
\item For both thin  and thick cases, acceptable fits in terms of reduced $\chi^2$, $\chi^2_\nu$,  and normalised residuals
were found for all simulated spectra using expressions (\ref{bhI})
and (\ref{ThickIBH}), respectively. The fitted parameters were close
to the input ones - within $\sim10\%$ or less.

Concerning the computational time (see Table 1) for the fitting
procedures, the SSW thin fit routine was about 2 times slower than
our method for similar accuracy while the SSW thick fit routine is
10-20 times slower than our method.

\item As regards the matter of trying to get a meaningful fit to the actual form
of $I(\epsilon)$ from truncated power-laws
$F(E),{\cal F}_0(E_0)$ by using double power-law fits to the photon
spectra (constant $\gamma$ above and below some break energy
$\epsilon_b$ - \verb(bpow.pro() we found that this failed to produce an acceptable
overall fit (generally, $\chi^2_\nu>2$) and that normalised
residuals clustered near $\epsilon\sim E_1,E_{01}$. Using a fixed value of
$\gamma_1$ such as $1$, $1.5$, $1.7$ did not help.

\end{enumerate}

\begin{table}
\begin{tabular}{ccccc}
 & SSW-thick & Eq. (\ref{ThickIBH}) & SSW-thin & Eq. (\ref{bhI}) \\
\hline \hline
$\delta,\delta_0=3$\\$E_1,E_{01}=10$ & $7.45$ & $0.721$ & $1.27$ & $0.726$ \\
\hline
$\delta,\delta_0=3$\\$E_1,E_{01}=20$ & $7.32$ & $0.594$ & $0.951$ & $0.726$ \\
\hline
$\delta,\delta_0=3$\\$E_1,E_{01}=30$ & $9.99$ & $0.585$ & $0.931$ & $0.599$ \\
\hline
$\delta,\delta_0=5$\\$E_1,E_{01}=10$  & $9.99$ & $0.721$ & $1.36$ &   $0.724$ \\
\hline
$\delta,\delta_0=5$\\$E_1,E_{01}=20$ &  $7.46$ & $0.594$ & $1.07$ &  $0.595$ \\
\hline
$\delta,\delta_0=5$\\$E_1,E_{01}=30$ & $7.05$  & $0.583$ & $1.20$ & $0.597$ \\
\hline \hline
\end{tabular}
\caption{Computational cost test. The test was performed on a 3.2 GHz PC with 1GB RAM.
Times are in seconds.}
\end{table}

Adding a plausible isothermal component with the above values for
parameters $EM$ and $T$ to the thin and thick target spectra
modifies the fit behaviour described above {\it only if} the thermal
component contributes significantly to or dominates the spectrum at
$\epsilon \ga E_1,E_{01}$. For such spectra, e.g. thin-target case $EM=1\times 10^{49}$ cm$^{-3}$, $kT=1.5$ keV, $\delta=3$, $E_1=10$~keV, the fitting functions introduced in this paper do give an acceptable fit but
only an upper limit on $E_1$ can be obtained. This limit is close
to the energy where the thermal spectrum steepens and falls below
the non-thermal part, fits with smaller $E_1$ being also consistent with the data since lost in the dominant thermal emission.

\begin{figure}
\includegraphics[width=80mm]{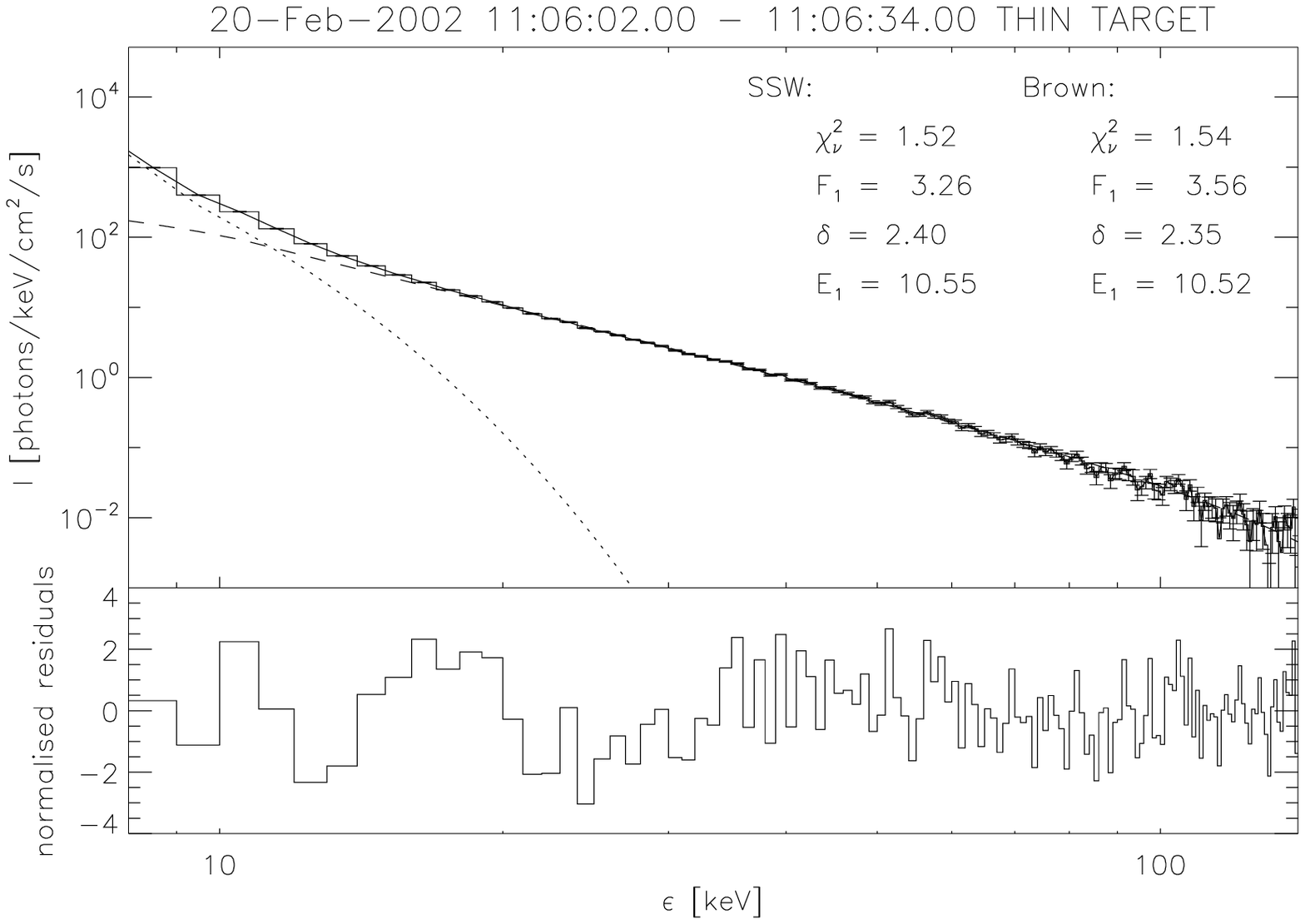}
\includegraphics[width=80mm]{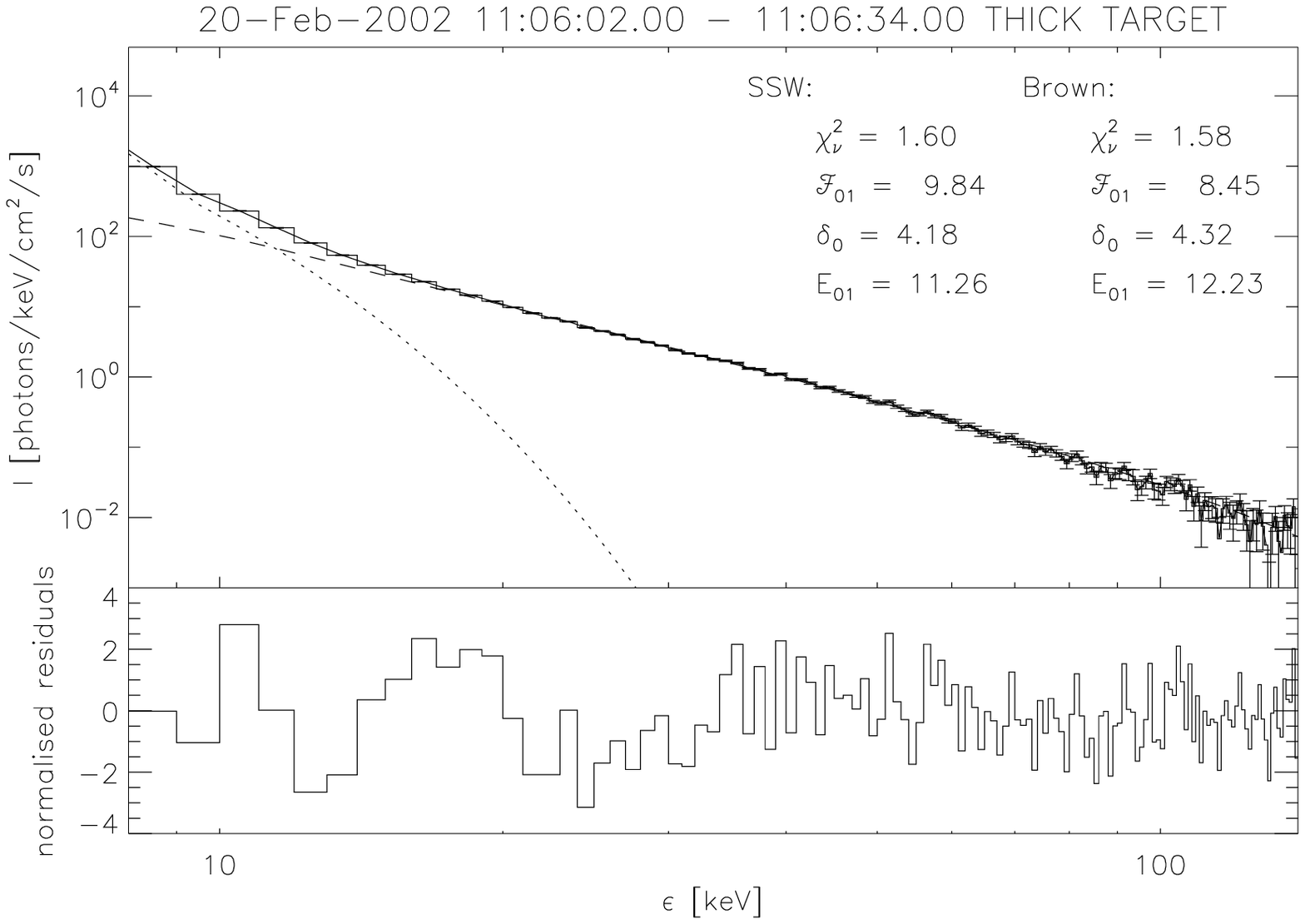}
\caption{Thin and thick target fits to February 20 2002 flare spectrum. The labels show the comparison between SSW and our approach.}
\end{figure}

\begin{figure}
\includegraphics[width=80mm]{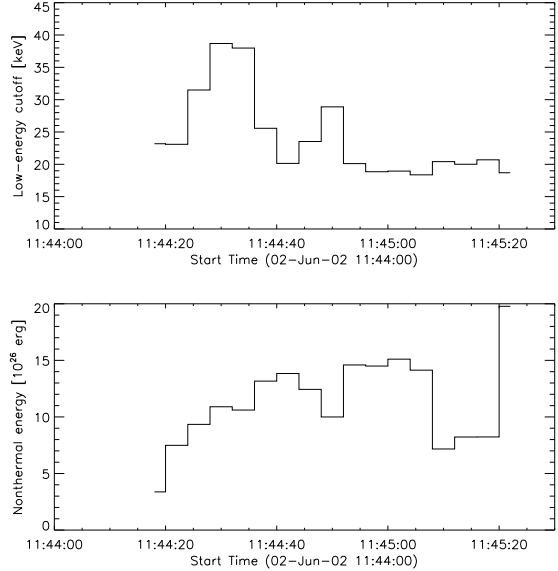}
\caption{Low-energy cutoff and total non-thermal electron energy in
4s time intervals for the June 2 2002 flare as derived from our
approach, see also Figure 4 in \citet{2007ApJ...670..862S}.}
\end{figure}

Photon spectra with a thermal component can also usually be fitted
with a nonthermal component close to a single power-law so the value
of $\epsilon_b$ obtained by double photon power law fits is not a
good indicator of $E_1,E_{01}$. Depending on the combination of thermal and
non-thermal parts, acceptable fits using \verb(bpow( yield
$\epsilon_b$ which can be either lower or higher than the input
$E_1,E_{01}$. Therefore, using $\epsilon_b$ for an estimate of the
non-thermal energy \citep[e.g.][]{2007arXiv0712.2544H} can be misleading. On
the other hand, our expressions generally give $E_1,E_{01}$ much closer to
the input $E_1,E_{01}$ than  $\epsilon_b$ is.

The proposed expressions have been also tested on two cases of real
data. Figure 2 shows thin and thick fits to the 20-Feb-2002 11 UT
flare and compares the parameters obtained by our and SSW
expressions. Both thin and thick fits give similar
${\overline{F}}(E)$ and ${\cal{F}}_0(E_0)$. This flare was near the
limb so required no albedo correction.

Next, we applied our thick target expression to the early impulsive
phase of the flare of 02-Jun-2002 which shows flattening and
evidence of a low-energy cutoff at $E_1$ above the thermal component
in the 18~--~38~keV range \citep{2007ApJ...670..862S}. In this case
albedo correction was applied, as is essential for such events.
Figure 3 shows the time evolution $E_1(t)$ obtained for the
time variation of the best fit low-energy cutoff and for the
corresponding total non-thermal electron power (as total energy per
 4 sec integration). These curves are closely comparable with those found by
\citet{2007ApJ...670..862S} in their Figure 4.

Thin and thick target formulas (Equations (\ref{bhI},\ref{ThickIBH}))
have been incorporated into the SSW tree. Prospective users may
access them as OSPEX fitting functions named \verb(photon_thin.pro(
and \verb(photon_thick.pro(.

\section{Conclusions}
We have shown that results for thin and thick target bremsstrahlung
photon spectra $I(\epsilon)$ from power-law electron spectra with
constant index $\delta$ and low-energy spectral cutoff $E_1,E_{01}$
obtained using the Bethe Heitler cross-section are very close to
those from the exact cross-section, at photon energies both above
and below the cutoff, at least at keV to deka-keV energies. We have shown
further that the Bethe Heitler expressions allow
the bremsstrahlung integrals for $I(\epsilon)$ to be written as
analytic forms in terms of beta functions of $\delta$
and $a$ only, which are part of standard numerical
packages, and that these give results very close to the exact
$I(\epsilon)$.

For both the thin and especially for the thick target models, we
find that this formulation enables equally accurate but much faster
spectral fitting than evaluation of the full spectral integrations
which will be valuable in analysis of bremsstrahlung HXR data such
as from RHESSI. This fast fit approach can also replace the use of
fitting unphysical broken power-laws in $I(\epsilon)$.

\acknowledgements{We kindly acknowledge financial support from ISSI,
UK STFC and UC Berkeley SSL (JCB), the Italian MIUR (AMM and MP) and
grant 205/06/P135 of the Grant Agency of the Czech Republic (JK). The
paper has benefited from discussions with H.S. Hudson, A.Caspi and
P.Saint-Hilaire.}
$$ $$

%\begin{thebibliography}{}

\bibliographystyle{aa}
\bibliography{refs_aa}
%\bibitem{br73}Brown, J.C. 1973, Sol. Phys.29,421
%
%\bibitem{}Brown, J.C. 2005, Solar  Magnetic Phenomena, Proceedings of
%the 3rd Summerschool and Workshop held  at the Solar Observatory
%Kanzelh" ohe, K" arnten, Austria, August 25  -- September 5, 2003.
%Edited by A.Hanslmeier, A.Veronig, and  M.Messerotti. Astronomy and
%Astrophysics Space Science Library, vol.320,  Springer, Dordrecht,
%The Netherlands, 2005., p.87-114, 87
%

%\bibitem{bretal06}Brown, J.C., Emslie, A.G., Holman, G.E., Johns, C.J., Kontar,
%E.P., Massone, A.M.,\& Piana, M. 2006, ApJ, 643, 523
%
%\bibitem{em78}Emslie, A.G. 1978, ApJ
%
%\bibitem{haetal07} Hannah, I. G., Hurford, G. J., Hudson, H. S., Lin, R. P. \& van Bibber, K. 2007, ApJ, 659,277
%
%\bibitem{haetal08} Hannah, I. G., Christe, S., Krucker, S, Hurford, G. J., Hudson, H. S. \& Lin, R. P. 2008, ApJ, in press
%
%
%\bibitem{}Holman G.D. 2003, ApJ 586, 606
%
%\bibitem{} Hudson HS ???? re constant gamma error
%
%\bibitem{} Other AN ????  re constant gamma error
%
%  Kontar et al 1995
%

%%
%\bibitem{lietal02} Lin, R. P.; Dennis, B. R.; Hurford, G. J.; Smith, D. M.; Zehnder,
%A.; Harvey, P. R.; Curtis, D. W.; Pankow, D.; Turin, P.; Bester, M.;
%and 56 coauthors : 2002 Solar Phys.210, 3

%\bibitem{}Piana M. , Barrett R.K., Brown J.C., and McIntosh S.W. : 2000,
%Inverse Problems  15, 1469

%
%\bibitem{suhude07} Sui, L., Hurford, G. D. \& Dennis, B. R. 2007, ApJ, 670, 862
%
%\bibitem{}Thompson A.M., Brown J.C., Craig, I.J.D. and Fülber, C.: l992,
%Astron. Astrophys. 265, 278-
%
%\end{thebibliography}

%\bibliography{report}
\end{document}